\documentclass[a4paper]{jpconf}
\usepackage[utf8]{inputenc}
\usepackage{graphicx}
\usepackage{amsmath,amssymb}
\usepackage{amsfonts}
\usepackage{physics}
\usepackage{xspace}
\usepackage{bm}
\usepackage{mathrsfs}
\usepackage{nicefrac}
\usepackage[dvipsnames]{xcolor}
\usepackage[unicode]{hyperref}
\hypersetup{colorlinks=true, citecolor=MidnightBlue, linkcolor=Blue, urlcolor=CornflowerBlue, linktocpage=true}
\usepackage[all]{hypcap}
\usepackage{multirow}
\usepackage[format=plain,justification=RaggedRight,labelsep=endash,font=small,labelfont=sc]{caption}
\usepackage[export]{adjustbox}

\definecolor {darkgreen}{rgb}{0.2,0.7,0.2}
\newcommand{\pdagger}{{\phantom{\dagger}}}
 
\begin{document}
\title{A compact formula for the quantum fluctuations of energy}
\author{ Rajeev Singh     }
\address{ Institute  of  Nuclear  Physics  Polish  Academy  of  Sciences,  PL-31-342  Krak\'ow,  Poland}
\begin{abstract}
A formula to calculate the quantum fluctuations of energy in small subsystems of a hot and relativistic gas is derived. We find an increase in fluctuations for subsystems of small sizes, but we agrees with the energy fluctuations in the canonical ensemble if the size is large enough. Not only one can use our expression to find the limit of the concepts of energy density or fluid element in connection to relativistic heavy-ion collisions, but also in other areas of physics where one studies matter with high temperature and velocity.
\end{abstract}
\section{Introduction}
\smallskip
Quantum fluctuations of physical observables, which commonly arise from quantum nature of the system~\cite{Huang:1987asp}, play a crucial role because they give out information about phase transitions~\cite{Huang:1987asp,Smoluchowski,PhysRevLett.85.2076,Gross:1980br,Haussler:2008gx,Herzog:2016kno,Vovchenko:2016dix,Steinheimer:2016cir,Lohr-Robles:2021ihl,Bai:2021wrh,Fortunato:2021fby,Li:2021qcb,Yang:2021hmg,Sami:2021ufn,deBrito:2021pmw}, and large scale structure formation~\cite{Lifshitz:1963ps,PhysRevLett.49.1110,Choudhury:2013woa,Choudhury:2016cso,Choudhury:2016pfr,Choudhury:2021tuu,Graef:2021tid} as well as phenomena which are dissipative in nature~\cite{Kubo1,Berges:1998rc,Halasz:1998qr,Stephanov:1998dy,Stephanov:1999zu,Hatta:2002sj,Son:2004iv,Stephanov:2008qz,Berdnikov:1999ph,CaronHuot:2011dr,Kitazawa:2013bta,Goswami:2020yez,Fu:2021wyc,Pradeep:2021opj,Goswami:2021opr,Bollweg:2021vqf,Schmidt:2021pey,Guo:2021qtf}.
This work studies the quantum fluctuations of energy in small systems (in other words, subsystems) of a hot and relativistic gas~\cite{Das:2020ddr,Das:2021aar,Das:2021rck,Singh:2021tih,Singh:2021dbu}. These fluctuations increase for small size of the subsystem, but agrees with the thermal fluctuations in the canonical ensemble if the size of the subsystem is large enough. Our results can be very helpful in determining the size of the subsystem (for specific particle mass and temperature) for which quantum energy fluctuations reach classical limit and may be ignored in such studies. We are specifically interested in the hot matter system which is believed to be produced during the relativistic heavy-ion collisions, and relativistic hydrodynamics has became a widely accepted theory to describe the physics of heavy-ion collisions~\cite{Fukushima:2013rx,Jaiswal:2016hex,Florkowski:2017olj,Fukushima:2018grm,Romatschke:2017ejr,Bhadury:2021oat}.
Within the theory of relativistic hydrodynamics, we need to use the concepts of energy density and pressure in order to characterize the fluid cells locally, which give rise the question that how the definition of energy density of such a small system can be defined properly having the size of about 1~fm.
In this article, we calculate and derive a formula to calculate the quantum fluctuations and then use this expression for some particular physical situations in the context of relativistic heavy-ion collisions.
We use the metric convention as $g_{\mu\nu} = \hbox{diag}(+1,-1,-1,-1)$. Bold font is used to denote three-vectors and the scalar product of both four- and three-vectors is represented by a dot, i.e., $a^\mu b_\mu = a \cdot b = a^0 b^0 - \boldsymbol{a} \cdot  \boldsymbol{b}$.
\smallskip
\section{Basic definitions}
\smallskip
We assume a subsystem $S_a$ of the larger
thermodynamic system $S_V$ which contain the spinless boson particles having mass $m$. This system $S_V$
is defined by the canonical ensemble and 
specified by the temperature $T$, where $\beta = 1/T$. Note that, volume $V$ of $S_V$ is larger than the volume of $S_a$, and $V$ is large enough for doing integrals over the momentum of the particles.
We characterize our bosonic system by a quantum scalar field which is in thermal equilibrium as
~\cite{Chen:2018cts}
\begin{equation}
\phi(t,{\boldsymbol{x}})=\int\frac{d^3k}{\sqrt{(2\pi)^3 ~2\omega_{\boldsymbol{k}}}}\left(a_{\boldsymbol{k}}^{\pdagger}e^{-i k \cdot x} +
a_{\boldsymbol{k}}^{\dagger}e^{i k \cdot x} \right),
\label{equ1ver1}
\end{equation}
with $a_{\boldsymbol{k}}^{\pdagger}$ and $a_{\boldsymbol{k}}^{\dagger}$ being the annihilation and creation operators, respectively, and $\omega_{\boldsymbol{k}}=\sqrt{{\boldsymbol{k}}^2+m^2}$ is the particle's energy. For the calculation of thermal averaging we need to
know the following expectation values of the products of two and four creation and/or annihilation operators~\cite{Itzykson:1980rh,Evans:1996bha}. Other combinations of creation and annihilation operators can be obtained easily from the following expression
%
\begin{align}
\langle a^{\dagger}_{{\boldsymbol{k}}}a_{{\boldsymbol{k}}^{\prime}}^{\pdagger}\rangle&=\delta^{(3)}({\boldsymbol{k}}-{\boldsymbol{k}}^{\prime})f(\omega_{{\boldsymbol{k}}}),\label{equ2ver1}\\
\langle a^{\dagger}_{{\boldsymbol{k}}}a^{\dagger}_{{\boldsymbol{k}}^{\prime}}a_{{\boldsymbol{p}}}^{\pdagger}a_{{\boldsymbol{p}}^{\prime}}^{\pdagger}\rangle &= \bigg(\delta^{(3)}({\boldsymbol{k}}-{\boldsymbol{p}})~\delta^{(3)}({\boldsymbol{k}}^{\prime}-{\boldsymbol{p}}^{\prime}) +\delta^{(3)}({\boldsymbol{k}}-{\boldsymbol{p^{\prime}}})~\delta^{(3)}({\boldsymbol{k}}^{\prime}-{\boldsymbol{p}})\bigg)f(\omega_{{\boldsymbol{k}}})f(\omega_{{\boldsymbol{k}}^{\prime}}).
\label{equ3ver1}
\end{align}
where $f(\omega_{{\boldsymbol{k}}})$ is the Bose--Einstein distribution function, $f(\omega_{{\boldsymbol{k}}})=1/(\exp[\beta ~\omega_{{\boldsymbol{k}}}]-1)$.

Now we define~\cite{Chen:2018cts} the following operator $\mathcal{H}_a$ which characterizes the energy density of a {\it finite} size subsystem $S_a$ which is placed at the origin of the coordinate system where $\mathcal{H}(x) \equiv (\dot{\phi}^2+({\bf{\nabla}\phi)}^2+m^2\phi^2)/2$ is the Hamiltonian density for the free real scalar field
\begin{align}
\mathcal{H}_a = \frac{1}{(a\sqrt{\pi})^3}\int d^3{\boldsymbol{x}}~\mathcal{H}(x)~\exp\left(-\frac{{\boldsymbol{x}}^2}{a^2}\right).
\label{equ4ver1}
\end{align}
We used above the Gaussian profile in order to remove boundary effects which might come from the sharp boundaries. 

Now one can easily obtain the thermal average of $\mathcal{H}_a$ as~\cite{Huang:1987asp}
\begin{align}
    \langle :\mathcal{H}_a :\rangle= \int \frac{d^3{{k}}}{(2\pi)^3}~\omega_{{\boldsymbol{k}}}~f\left(\omega_{{\boldsymbol{k}}}\right) \equiv \varepsilon(T).
    \label{equ5ver1}
\end{align}
To arrive at the Eq.~\eqref{equ5ver1}, we used normal ordering in order to remove the infinite vacuum contribution coming from the zero-point energy term. Eq.~\eqref{equ5ver1} is also independent of the size $a$ reflecting the spatial uniformity of the thermodynamic system $S_V$.
In order to calculate the quantum fluctuation of energy of the subsystem $S_a$
we use the formula for variation and the normalized standard deviation, given respectively as
\begin{align}
 \sigma^2(a,m,T) &= \langle :\mathcal{H}_a: :\mathcal{H}_a: \rangle - \langle :\mathcal{H}_a :\rangle^2\, ,
 \label{sigma2}\\
\sigma_n(a,m,T) &= \frac{(\langle:\mathcal{H}_a::\mathcal{H}_a:\rangle- \langle :\mathcal{H}_a :\rangle^2)^{1/2}}{\langle :\mathcal{H}_a :\rangle}\, .
 \end{align}
Using Eqs.~\eqref{equ2ver1} and \eqref{equ3ver1},  we find
\begin{align}
\sigma^2(a,m,T) &=  \int dK ~dK^{\prime} f(\omega_{{\boldsymbol{k}}})(1+f(\omega_{{\boldsymbol{k}}^{\prime}}))
\bigg[(\omega_{{\boldsymbol{k}}}\omega_{{\boldsymbol{k}}^{\prime}}+{\boldsymbol{k}}\cdot{\boldsymbol{k}}^{\prime}+m^2)^2e^{-\frac{a^2}{2}({\boldsymbol{k}}-{\boldsymbol{k}}^{\prime})^2}\nonumber\\
&+(\omega_{{\boldsymbol{k}}}\omega_{{\boldsymbol{k}}^{\prime}}+{\boldsymbol{k}}\cdot{\boldsymbol{k}}^{\prime}-m^2)^2e^{-\frac{a^2}{2}({\boldsymbol{k}}+{\boldsymbol{k}}^{\prime})^2}\bigg].
\label{equ6ver1}
\end{align}
where $dK = d^3{{k}}/((2\pi)^{3} 2 \omega_{{\boldsymbol{k}}})$ and neglected a temperature independent divergent term which may be related to the pure vacuum energy fluctuation~\cite{Phillips:2000jm}. 

Eq.~\eqref{equ6ver1} is the main result of ours which determine the quantum fluctuations of energy of the ``Gaussian'' subsystems $S_a$ of
the larger thermodynamic system $S_V$.
Numerically we can calculate the results for any system of size $a$, temperature $T$, and
mass $m$, the plots of which will be shown below.

One should also keep in mind the factor of degeneracy which is important when we study the thermodynamic properties of particles. This degeneracy factor is related to the internal degrees of freedom, for instance, spin, isospin or color charge. Therefore to take into consideration in our framework, the $g$ copies of particles, one needs to consider the $g$ copies of scalar field which is given by both creation and annihilation operators which commute for different species of particles. Hence, the following replacements are required
for the calculation of the quantum fluctuations
\begin{equation}
\varepsilon \rightarrow g \varepsilon, \quad
\sigma^2 \rightarrow g \sigma^2.
\label{g}
\end{equation}
%
\section{Thermodynamic limit}
\smallskip
We believe that in the limit of large system size, that is, 
$a\rightarrow \infty$
(but still with $a^3 \ll V$),
our quantum fluctuation expression should agree with classical statistical mechanics~\cite{Huang:1987asp}. Hence, we use the following representation of the Dirac delta function in Eq.~\eqref{equ6ver1}
\begin{align}
    \delta^{(3)}({\boldsymbol{k}}-{\boldsymbol{p}})=\lim_{a \to\infty} \frac{a^3}{(2\pi)^{3/2}}e^{-\frac{a^2}{2}({\boldsymbol{k}}-{\boldsymbol{p}})^2}.
\end{align}
which give rise to the formula valid in the large size $a$ limit
\begin{align}
\sigma^2 & \sim 
\frac{g}{(2\pi)^{3/2} a^3}
\int \frac{d^3{{k}}}{(2\pi)^3}~\omega_{{\boldsymbol{k}}}^2~f(\omega_{{\boldsymbol{k}}}) (1+f(\omega_{{\boldsymbol{k}}})).
\nonumber
\end{align}
Note that we calculate the quantum fluctuation for massless particles as $\sigma^2 \sim T^5/a^3$, therefore RHS of the last equation can be written as
\begin{equation} 
c_V = \frac{d\varepsilon}{dT} = \frac{g}{T^2} \int \frac{d^3{{k}}}{(2\pi)^3}~\omega_{{\boldsymbol{k}}}^2~f(\omega_{{\boldsymbol{k}}}) (1+f(\omega_{{\boldsymbol{k}}})).
\end{equation}
where $c_V$ is  
the specific heat at constant volume.
Hence, in large $a$ limit~\cite{Mrowczynski:1997mj}
\begin{align}
V_a \sigma_n^2
= \frac{T^2 c_V}{\varepsilon^2}
= V \frac{\langle H^2\rangle-\langle H \rangle^2}{\langle H \rangle^2} \equiv V \sigma^2_H,
\label{equ10ver1}
\end{align}
with $V_a = a^3 (2\pi)^{3/2}$ and $H$
being the Hamiltonian of $S_V$,
and RHS of Eq.~\eqref{equ10ver1}
is the normalized energy fluctuation in the thermodynamical system $S_V$~\cite{Huang:1987asp}.
Here, $V_a=a^3 (2\pi)^{3/2}$
is the volume of $S_a$.
\begin{figure}[t]
\begin{center}
	\includegraphics[scale=0.45]{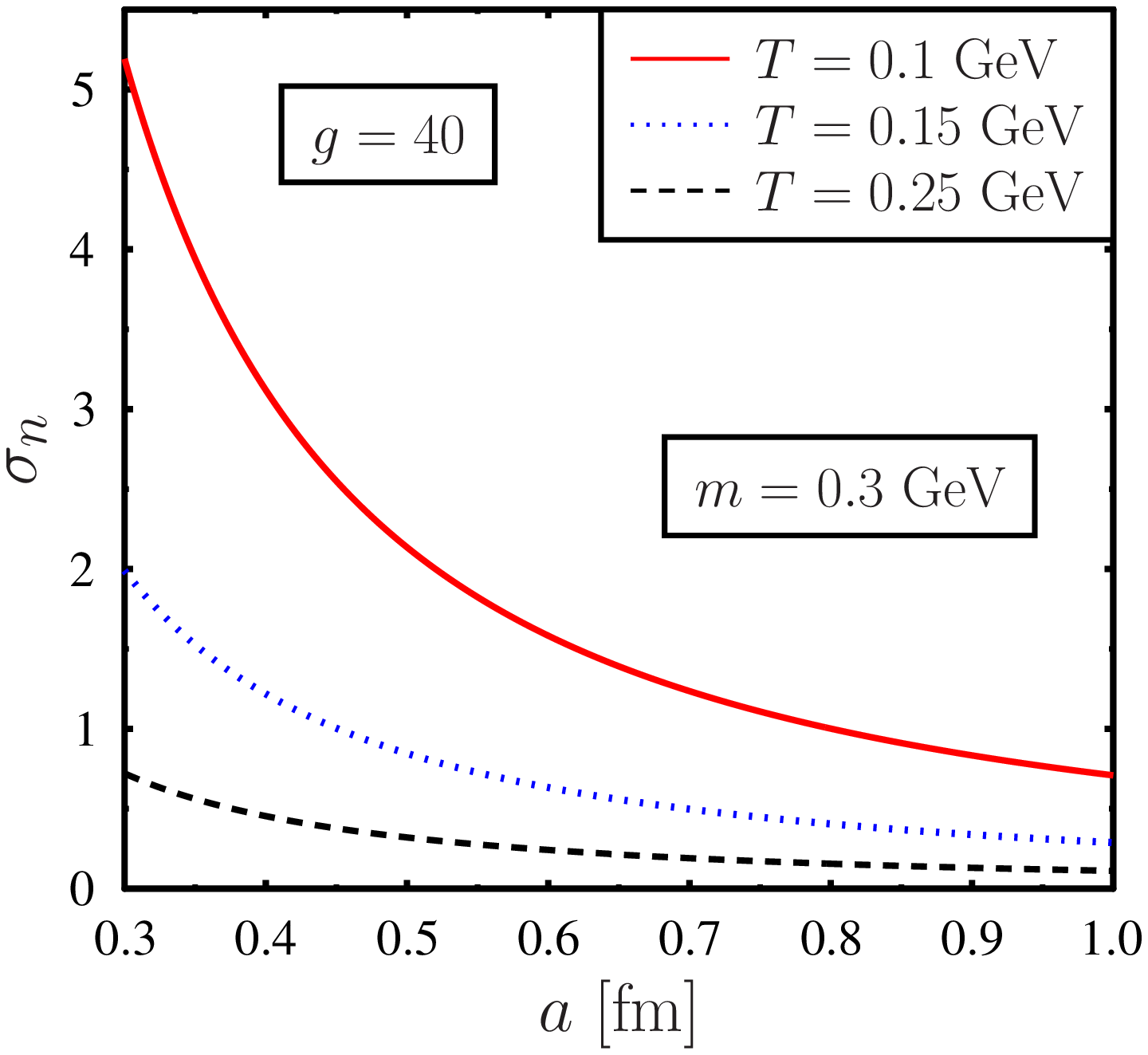}
	\includegraphics[scale=0.45]{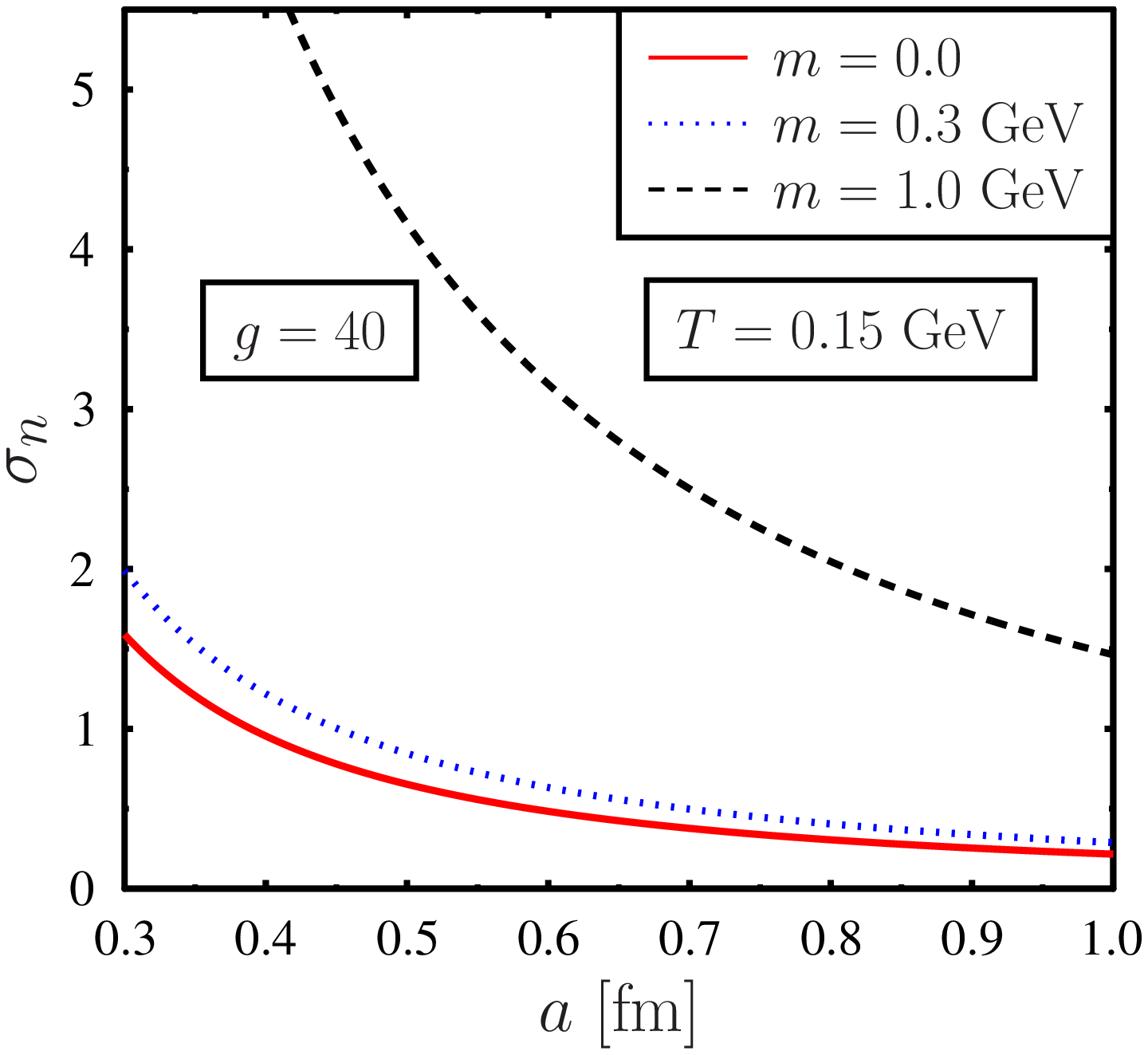}
	\caption{(up) Variation of the normalized energy density fluctuation $\sigma_n$ in the subsystem $S_a$ with size $a$ for various values of the temperature $T$ but with fixed particle mass $m=0.3$ GeV. (down) Same as above figure but for fixed temperature $T=0.15$~GeV and different particles masses.}
	\label{fig:m_0.3}
\end{center}
\end{figure}
\section{Numerical results} 
\smallskip
Since we correctly reproduced thermodynamic limit for $a\rightarrow\infty$, we can do some numerical calculations. 
Fig.~\eqref{fig:m_0.3} show the variation of the normalized energy density fluctuation $\sigma_n$ having the size $a$ for
the subsystem with various values of temperature and mass of the system.
We use degeneracy factor to be
$g=40$
in order to be consistent with the heavy-ion experiments~\cite{Kisiel:2005hn}.
Fig.~\eqref{fig:m_0.3} indicate that $\sigma_n$ decreases with increasing the system size
$a$
as expected from the behavior of fluctuations. Since, from Eq.~\eqref{equ6ver1} we see that $\sigma_n$ explodes in the limit $a \to 0$ hence we start our numerical calculations at $a=0.3$~fm.
Having fixed mass of the particle we find the decrease in the normalized fluctuations with the increase in temperature. On the other hand, at fixed $T$ the fluctuations grow with $m$.
\begin{figure}[t]
\begin{center}
\includegraphics[scale=0.45]{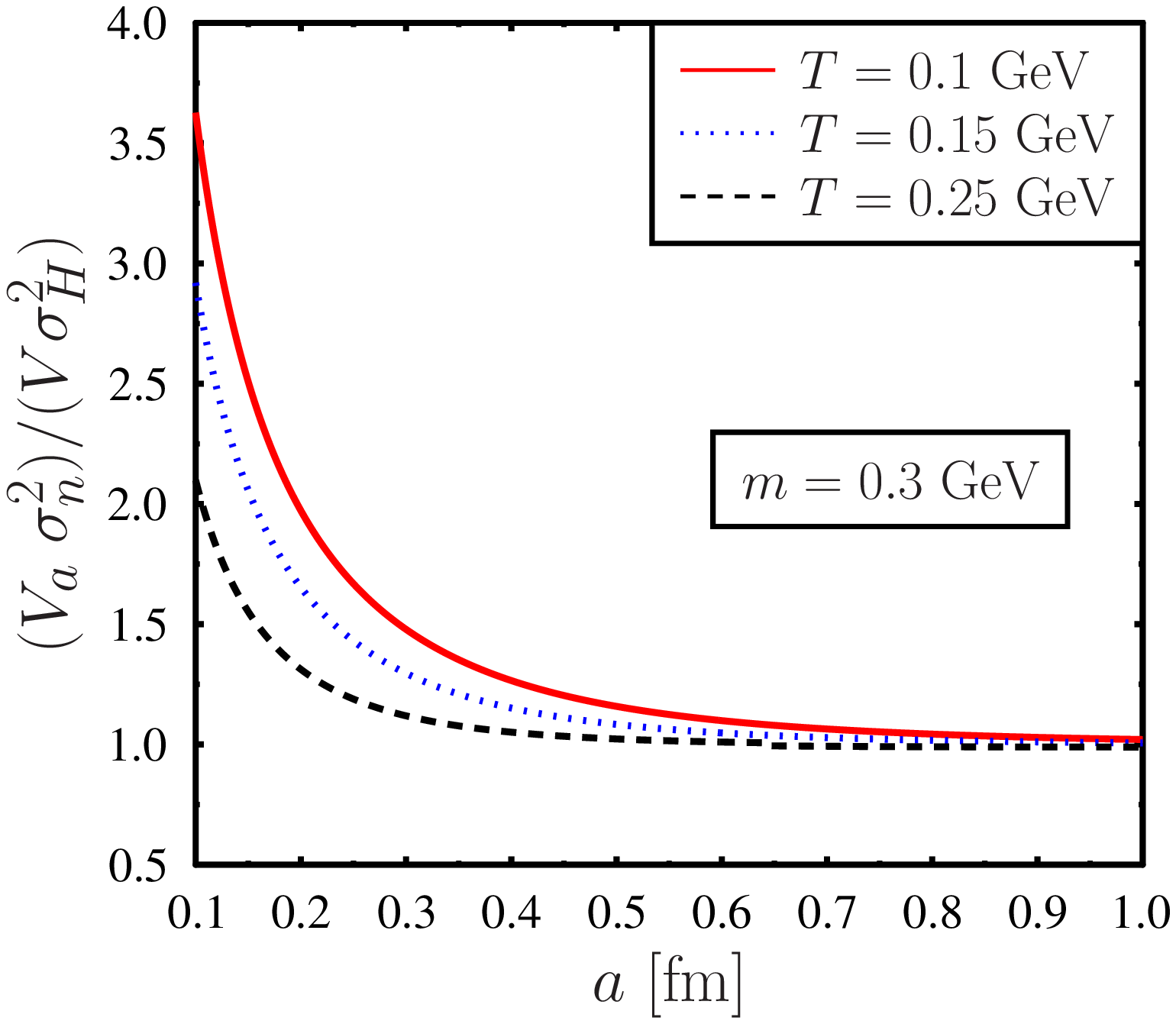}
\includegraphics[scale=0.45]{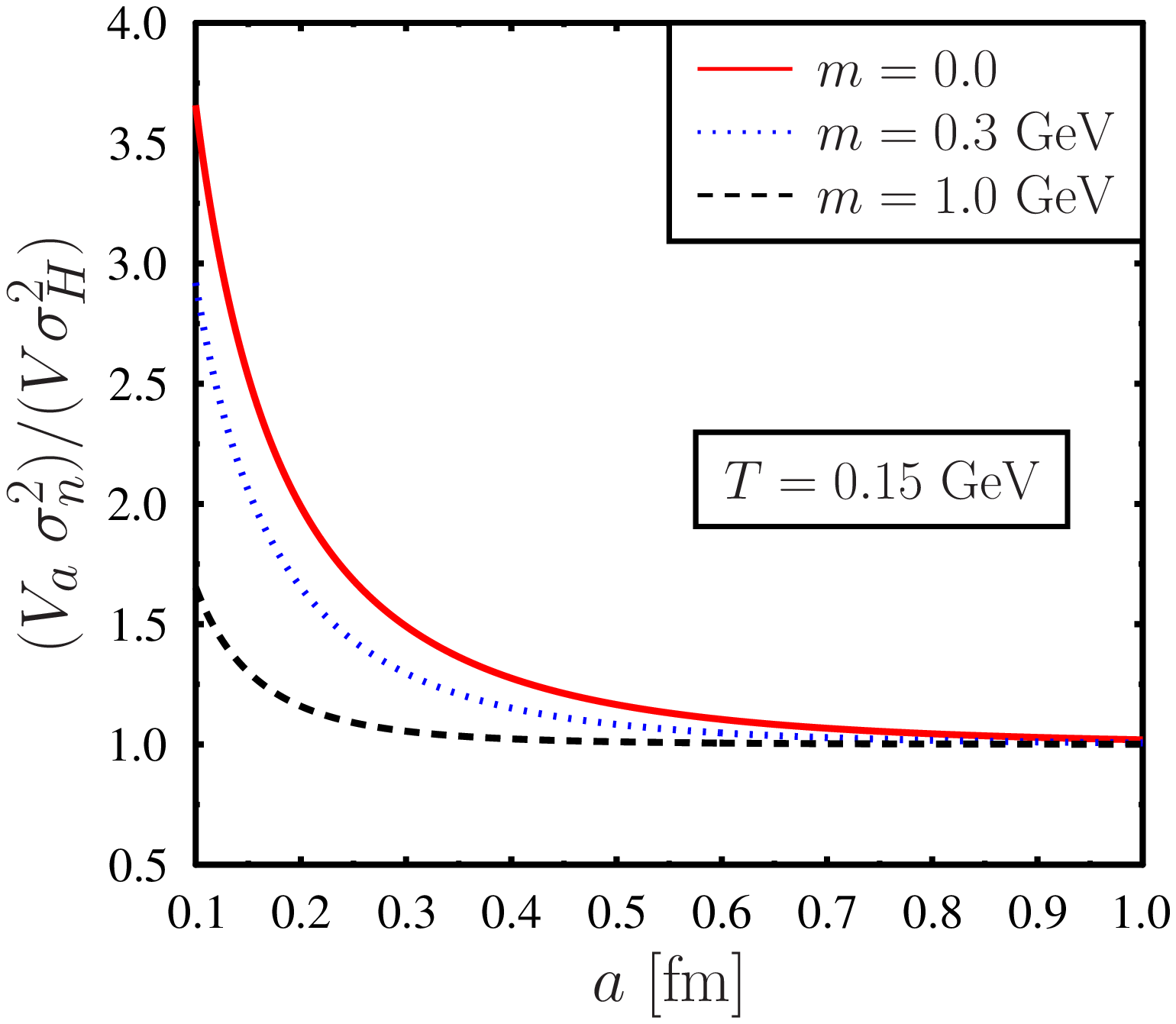}
\caption{(up) Variation of the normalized energy fluctuation in the subsystem $S_a$ with the size $a$ for various values of the temperature at $m=0.3$ GeV.
(down) Same as above figure but for different values of mass at $T=0.15$ GeV.
}
\label{fig:norm_0.3}
\end{center}
\end{figure}

Fig.~\eqref{fig:norm_0.3} presents the variation of $V_a \sigma_n^2/V\sigma^2_H$ with the subsystem size $a$
for particles with mass $m$
obeying Bose-Einstein statistics.
From Eq.~\eqref{equ10ver1} we see that it approaches the thermodynamic limit~\cite{Kapusta:2011gt} and hence $V_a\sigma_n^2/V\sigma^2_H$ should go to unity, which is seen from Fig.~\eqref{fig:norm_0.3}. 
\section{Conclusions} 
\smallskip
In this work we have derived the expression for the quantum fluctuation of energy for the subsystems of a hot and relativistic gas which beautifully agrees with the thermodynamic fluctuation expression for the large system size.
We also explained and outlined the possible consequences of our results for the description of relativistic heavy-ion systems.


\section*{Acknowledgments}
I am very grateful and thankful to A. Das, W. Florkowski, and R. Ryblewski for their beautiful collaboration. This research was supported in part by the Polish National Science Centre Grants No. 2016/23/B/ST2/00717 and No. 2018/30/E/ST2/00432, and IFJ PAN.
\section*{References}
\bibliographystyle{iopart-num}
\bibliography{fluctuationRef.bib}{}
\end{document}